\documentclass[aps,pra,twocolumn,showpacs,superscriptaddress,10pt]{revtex4-1}
\usepackage{amsmath,amssymb,graphicx,epsfig}
\begin{document}

\title{Noise conversion in Kerr comb RF photonic oscillators}

\author{Andrey B. Matsko}\email{Corresponding author: andrey.matsko@oewaves.com}
\author{Lute Maleki}
\affiliation{OEwaves Inc., 465 North Halstead Street, Suite 140, Pasadena, CA 91107}

\begin{abstract}
Transfer of amplitude and phase noise from a continuous wave optical pump to the repetition rate of a Kerr frequency comb is studied theoretically, with focus on generation of spectrally pure radio frequency
(RF) signals via demodulation of the frequency comb on a fast photodiode. It is shown that both the
high order chromatic dispersion of the resonator spectrum and frequency-dependent quality factor of
the resonator modes facilitate the optical-to-RF noise conversion that limits spectral purity of the
RF signal.
\end{abstract}

\maketitle

\section{Introduction}

In a radio frequency (RF) photonic oscillator that utilizes an optical frequency comb the signal is produced by demodulating the equally spaced frequency harmonics of the comb on a fast photodiode. The relative coherence, or phase locking of the optical frequency
harmonics, is a necessary condition for generation of spectrally pure RF signals. The performance of
the best photonic generators is far superior to electronic devices. For instance, generation of 10~GHz signals
with phase noise at the level of -100 dBc/Hz at 1~Hz offset and -179~dBc/Hz at 1~MHz and higher
offsets were demonstrated with a femtoscond mode locked laser \cite{fortier11np,quinlan13np}.
This phase noise level is 40~dB lower  compared with phase noise of the best room-temperature
electronic oscillator. The achieved spectral purity of the photonic device could be spoiled by back-end
electronic amplifiers because of their inherent noise. However, another attribute of the photonic oscillator is that amplifiers are usually not
required, since the output power of a 10~GHz RF  source can reach 25~mW when a uni-traveling
carrier photodiode is utilized \cite{fortier13ol}.

Phase noise is the main characteristic determining the performance of an oscillator and its practical
usefulness. Optical monolithic microresonator-based RF photonic oscillators are advantageous for
providing low noise RF signals on a chip.  Development of RF generation with hyper-parametric
oscillators and mode-locked (Kerr) frequency combs has recently become the major direction in this area
\cite{savchenkov04prl,savchenkov08prl,savchenkov08oe,papp11pra,delhaye11prl,savchenkov13ol}. For
example, phase noise of a whispering gallery mode (WGM) resonator-based hyper-parametric RF photonic
oscillator is much lower as compared with the noise of existing electronic oscillators such as
dielectric resonator oscillators. Packaged 35~GHz Kerr comb RF photonic oscillators with phase noise
of -115 dBc/Hz at 10~kHz offset and -130 dBc/Hz at 100~kHz offset were previously demonstrated and have
became commercially available \cite{savchenkov13ol}. A miniature 10~GHz RF photonic oscillator
characterized with phase noise better than -60~dBc/Hz at 10~Hz, -90~dBc/Hz at 100~Hz, -120~dBc/Hz at
1~kHz, and -150~dBc/Hz at 10~MHz was demonstrated recently \cite{liang14spie}. The frequency stability
of this device is at the level of $10^{-10}$ at 1-100~s integration time \cite{liang15tbp}, which is
better than existing RF photonic devices of similar size, weight, and power consumption.

Quantum noise limited phase diffusion of the microresonator-based hyper-parametric RF photonic
oscillator can be described by a Schawlow-Townes-like formula \cite{matsko05pra}. A Leeson model can
be used far enough from the RF carrier to understand the behavior of the phase noise
\cite{savchenkov08oe}. The phase noise of the Kerr comb oscillator operating in the mode locked
regime has been recently analyzed as well \cite{matsko13oe}. In accordance with the obtained theoretical formula, the
phase noise of the mode locked Kerr frequency comb oscillator made with a 7~mm magnesium fluoride
WGM resonator, having a quality factor of $10^9$ and pumped with 5~mW of optical power,  is expected to be -90~dBc/Hz at 10~Hz, -110~dBc/Hz at 100~Hz, -130~dBc/Hz at
1~kHz, and -154~dBc/Hz at 10~MHz. The predicted noise floor, determined by the shot noise of the
oscillator, matches the measured value \cite{liang14spie} well. The measured close-in noise, however, is significantly worse than the predicted number. Evidently, there exists a mechanism that increases the oscillator noise well
above the quantum limit.

\begin{figure}[ht]
  \centering
  \includegraphics[width=6.5cm]{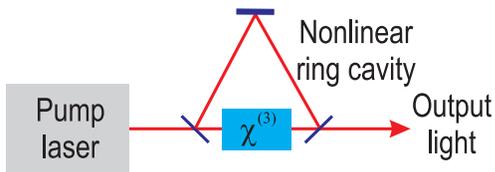}
\caption{ \small  Schematic diagram of a Kerr frequency comb generator. A continuous wave coherent light source pumps a mode of a nonlinear ring resonator. Optical harmonics are produced in the resonator modes if the pump power exceeds  a certain threshold. A mode locked operation of the device is possible, in which the frequency harmonics correspond to a train of short optical pulses in time domain. } \label{fig1}
\end{figure}

In this paper we analyze classical transfer functions describing the impact of  amplitude and frequency noise of the pump laser on the repetition rate of the Kerr frequency comb generated in a continuously pumped monolithic microresonator (see Fig.~\ref{fig1} as an example of the frequency comb generator). Using this result, we analyze the noise of the Kerr comb RF photonic oscillator and show that classical noise of the pump laser can be an important limiting factor on the performance of the oscillator.

We start with a simple analytical model of the resonant hyper-parametric oscillation that involves interaction of light confined in three resonator modes. The central mode is pumped optically. The oscillation can be described as a nondegenerate parametric process in which two pump photons are transformed to two different photons of lower and higher frequencies. This occurs when the pump power exceeds a certain threshold determined by the loss of the system, and by nonlinear coupling efficiency between the resonator modes. We show that the beat frequency between the pump laser and generated sidebands does not depend on the laser  parameters, if all the modes have the same quality (Q-) factor. The dependence on the parameters of the pump light emerges as soon as the degeneracy of the Q-factor is lifted. Both amplitude and phase modulation of the pump can be transferred to the beat frequency of the optical harmonics of the oscillator. The phase and amplitude noise of the pump can impact the beat frequency as well.

A large variety of nonlinear processes takes place in an optically pumped nonlinear microresonator. We here focus at mode-locked Kerr frequency comb and mode-locked hyper-parametric oscillation that behave in a similar way. Both processes are characterized with hard  excitation regime and the fundamental noise of both processes can be described by similar analytical expressions \cite{matsko13oe}. Generation of the mode-locked Kerr frequency comb is different from the hyper-parametric oscillation only in a sense that a large number of nearly equal power frequency harmonics are generated from the cw optical pump, in contrast with a pair of optical harmonics considered in the hyper-parametric oscillator model.

Mode-locked hyper-parametric oscillation occurs in resonators with relatively large group velocity dispersion, while Kerr combs are generated in resonators with a small dispersion value. To modify the dispersion value one needs to change loading of the resonator modes. It is possible to transfer a narrow frequency comb to a broad frequency comb adiabatically by changing the loading accompanied by proper change of the pump power and frequency. This fact proves physical identity of the combs.

Even though the narrow and broad combs are identical, an analytic description of noise transfer for the broad frequency comb is not straightforward. We perform a numerical simulation of the system and solve a set of ordinary differential equations describing the comb \cite{chembo10pra,matsko12pra} to find the impact of noise transfer process on a realistic Kerr comb oscillator. We evaluate the corresponding modulation of the comb power and repetition rate by adding modulation  to the amplitude and phase of the pump, and subsequently infer the corresponding transfer functions that aid estimating the impact of the pump noise on those parameters. There is no transfer if the resonator spectrum is characterized with second order group velocity dispersion (GVD) only and all the modes have the same Q-factor. By substituting realistic asymmetry parameters to the model we find that the impact of a noisy pump light can be rather significant. A natural way to reduce this influence is to utilize a low noise pump laser.

This paper is organized as follows. In Section I we provide an analytical model of hyper-parametric oscillation and study the dependence of its frequency on parameters of the pump light. Numerical simulations involving Kerr frequency comb are described in Section II. We discuss how the approximations considered in the paper are related to realistic resonators in Section III. Section IV concludes the paper.

\section{Pump-dependent frequency of a hyper-parametric oscillator}

The theory of hyper parametric oscillation presented here is related to the theory of additive modulational instability in ring lasers \cite{haelterman92ol,coen97prl,coen01ol}, and in fiber-based optical parametric oscillators \cite{serkland99ol,sharping02ol,matos04ol,deng05ol}. In the case of high-Q WGM hyper-parametric oscillator the theory was developed in \cite{matsko05pra}. We adopt the model \cite{maleki10ifcs} to analyze the dependence of the frequency of a hyper-parametric oscillator on the frequency and power of the pump laser. We show that an ideal symmetry in the oscillation process results in complete suppression of the influence of the pump parameters on the frequency interval between the pump light and generated optical frequency harmonics. This frequency interval coincides with the frequency of the RF signal generated by demodulation of the frequency comb on a fast photodiode. We argue that this is not the case in practice, since modes of an optical resonator have slightly different Q-factors. For instance, this difference is fundamental in the case of an evanescently coupled whispering gallery mode resonator.

\subsection{Basic equations and their steady state solution}

We consider a hyper-parametric oscillation process in which two monochromatic optical pump photons transform into two sideband photons mediated by the  cubic nonlinearity of the  ring resonator's host material. The oscillation is described by the Kerr Hamiltonian
\begin{equation} \label{v}
V=-\hbar (g/2) :(a+b_++b_-+h.c.)^4:,
\end{equation}
where "$:\dots :$" stand for normal ordering, $a$, $b_+$, and $b_-$ are the annihilation operators for the optically pumped and sideband modes, respectively; $g$ is the coupling constant.

The coupling constant $g$ can be derived using definition of the nonlinear refractive index:
\begin{equation} \label{n}
n = n_0+n_2 I,
\end{equation}
where $I$ denotes the time-averaged intensity of the optical field within the resonator mode, and $n_0$ and $n_2$ are the linear and nonlinear refractive indices of the resonator host material. According to (\ref{v}), the power-dependent frequency shift of the optically pumped mode depends on the number of photons in the mode ($\langle a^\dag a \rangle $) as $\Delta \omega_0 = -g\langle a^\dag a \rangle $. On the other hand, in accordance with (\ref{n}), $\Delta \omega_0 = -\omega_0 n_2 I/n_0$ ($\omega_0$ is the frequency of the mode). Since $\hbar \omega_0 \langle a^\dag a \rangle = {\cal V} n_0 I /c$ (${\cal V}$ is the mode volume), we get
\begin{equation}\label{g}
g= \omega_0 \frac{\hbar \omega_0 c}{{\cal V} n_0} \frac{n_2}{n_0}.
\end{equation}
This simple derivation is valid for the case of completely spatially overlapped resonator modes. The mode volume parameter has to be modified appropriately in a more general case.

Using Eq.~(\ref{v}) and applying the rotating wave approximation we derive equations describing the amplitude of the electric field within the optical modes
\begin{eqnarray} \label{olq1a}
\dot a &=& -(i \omega_0 (T)  + \gamma_0+\gamma_{c0}) a +
ig \bigl [ a^* a + 2 b_+^* b_+ + \\ \nonumber &&
2 b_-^* b_- \bigr ] a+
2iga^* b_+b_- +\sqrt{\frac{2\gamma_{c0} P_0}{\hbar
\omega_0}}\; e^{-i\omega t} ,
\\
\label{olq2a}  \dot b_+ &=& -(i \omega_+ (T) +\gamma_+ +
\gamma_{c+}) b_+ + \\ \nonumber &&  ig \bigl [ 2a^* a + b_+^* b_+ +  2
b_-^* b_- \bigr ] b_+ + igb_-^* a^2 ,
\\
\label{olq3a} \dot b_- &=& -(i \omega_- (T)   + \gamma_- +
\gamma_{c-}) b_- + \\ \nonumber &&  ig \bigl [ 2a^* a + 2 b_+^* b_+ +
b_-^* b_- \bigr ] b_- + igb_+^* a^2 ;
\end{eqnarray}
where $\omega_0(T)$, $\omega_+(T)$, and $\omega_-(T)$ are the temperature (T)-dependent eigenfrequencies of the WGMs;  $\gamma_0$, $\gamma_+$, and $\gamma_-$ are the internal decay rates of the modes; $\gamma_{c0}$, $\gamma_{c+}$, and $\gamma_{c-}$ are the decay rates due to external coupling; $P_0$ is the power of the external pump.

We are interested in a classical analysis of the oscillation, so we transfer the operators to the expectation values of field amplitudes. Additionally, to simplify the equations we introduce slowly varying amplitudes:
\begin{eqnarray} \label{amps}
a = A e^{-i\omega t}, b_+ = B_+e^{-i\tilde \omega_+ t}, \hat b_- =
B_- e^{-i\tilde \omega_- t},
\end{eqnarray}
where $\omega$ is the carrier frequency of the external pump, and $\tilde \omega_+$ and $\tilde \omega_-$ are the carrier frequencies of generated optical harmonics. Values of the frequencies of the harmonics are determined by the oscillation
process and cannot be controlled externally. However, from energy conservation principle we find a relationship between them:
\begin{equation} \label{del}
2 \omega = \tilde \omega_+ + \tilde \omega_-.
\end{equation}

We substitute (\ref{amps}) into (\ref{olq1a}-\ref{olq3a}) and derive a set of algebraic equations neglecting time derivatives of slowly varying amplitudes $A$ and $B_{\pm}$
\begin{eqnarray} \nonumber
 \Gamma_0   A   &=&  ig \bigl [  |  A  |^2
 + 2|   B_+   |^2 +
2 |   B_-   |^2\bigr ]   A   +  \\ && 2ig
A^*     B_+     B_-   +
F_{c0}  , \label{sseq1}
\\ \nonumber
\Gamma_{+}   B_+   &=&     ig (2|  A
 |^2+2|  B_-  |^2+|  B_+  |^2)
  B_+   + \\ &&  ig   B_-^*     A ^2,  \label{sseq2}
\\  \nonumber
 \Gamma_{-}   B_-   &=&    ig (2|  A
 |^2+|  B_-  |^2+2|  B_+
 |^2)  B_-   + \\ && ig   B_+^*
A ^2. \label{sseq3}
\end{eqnarray}
where
\begin{eqnarray} \nonumber
&& \Gamma_{0} =  i (\omega_0(T) -\omega) + \gamma_0 +
\gamma_{c0},
\\
\nonumber && \Gamma_{+} =  i (\omega_+ (T) - \tilde \omega_+) + \gamma_+ + \gamma_{c+},
\\
\nonumber && \Gamma_- = i (\omega_- (T) - \tilde \omega_- )
+ \gamma_- + \gamma_{c-},
\end{eqnarray}
are the complex parameters determining detuning and attenuation in the system.

To solve the equations it is convenient to introduce dimensionless parameters:
\begin{eqnarray}
&& \bar \gamma_j = \gamma_j+ \gamma_{cj}, \nonumber \\
&& \xi=\frac{g |  A  |^2}{\bar \gamma_0}, \nonumber \\
&& f = \left (\frac{g}{\bar \gamma_0} \right )^{1/2} \frac{|  F_{c0}   |}{\bar \gamma_0}, \nonumber \\
 &&  A   = |  A   | e^{i\phi_0}, \nonumber \\
 &&  B_\pm   = |  B_\pm   | e^{i \phi_\pm}, \nonumber \\
&& {\cal B}_\pm=\frac{|  B_\pm   |}{|  A   |}, \nonumber \\
&& \psi=\phi_{F_{c0}}-\phi_0, \nonumber \\
&&   F_{c0}  = |  F_{c0}   | e^{ i \phi_{F_{c0}}}, \nonumber \\
&& \phi= 2\phi_0-\phi_+-\phi_-, \nonumber \\
&& \Delta_\pm = \frac{\omega_\pm(T) - \tilde \omega_\pm }{\bar \gamma_0}, \nonumber \\
&& \Delta_0 =\frac{\omega_0(T)-\omega }{\bar \gamma_0}, \nonumber \\
&& D=\frac{2\omega_0(T) - \omega_+(T) - \omega_-(T)}{\bar \gamma_0} \simeq  \frac{\beta_2 c\omega_{FSR}^2}{\bar \gamma_0 n_0}. \label{dd}
\end{eqnarray}

It is easy now to transform the set of three equations (\ref{sseq1}-\ref{sseq3}) to the set of six algebraic equations using this parametrization
\begin{eqnarray} && \label{f1}
\sqrt{\xi}(1-2 \xi {\cal B}_+{\cal B}_- \sin \phi) = f \cos \psi,
\\ \label{f2} && \sqrt{\xi} \left \{ \Delta_0 - \xi
\left [ 1 + 2 ({\cal B}_+^2+{\cal B}_-^2+{\cal B}_+{\cal B}_- \cos
\phi) \right ] \right \}  \\ \nonumber && = f \sin \psi,
\\ \label{f3} && \bar \gamma_+ {\cal B}_++ \bar \gamma_0 \xi {\cal B}_- \sin \phi = 0,
\\ \label{f4} && \left [\Delta_+ - \xi (2+{\cal B}_+^2+2{\cal B}_-^2)\right ] {\cal B}_+ - \xi {\cal B}_- \cos \phi = 0,
\\ \label{f5} && \bar \gamma_- {\cal B}_-+ \bar \gamma_0 \xi {\cal B}_+ \sin \phi = 0,
\\ \label{f6}&& \left [ \Delta_- - \xi (2+2{\cal B}_+^2+{\cal B}_-^2) \right ] {\cal B}_- - \xi {\cal B}_+ \cos \phi = 0,
\\ \label{f7}&& \Delta_+ + \Delta_- = 2\Delta_0 - D \;\;\;(\tilde \omega_+ + \tilde \omega_-=2 \omega).
\end{eqnarray}
Parameters $\xi$, $\phi$, $\Delta_+$, $\Delta_-$ as well as ratio between ${\cal B}_+$ and ${\cal B_-}$ can be inferred from Eqs.~(\ref{f3}-\ref{f7}). They are
\begin{eqnarray} \label{xiamp}
&& \xi^2 = \frac{\bar \gamma_+ \bar \gamma_-}{\bar \gamma_0^2} +
\frac{\bar \gamma_+ \bar \gamma_-}{(\bar \gamma_+  +\bar
\gamma_-)^2} \\ \nonumber && \left [2 \Delta_0 - D-\xi (4 + 3({\cal B}_+^2 +{\cal
B}_-^2) )\right ]^2, \\ \label{sinphi}
&& \sin \phi = -\frac{\sqrt{\bar \gamma_+\bar \gamma_-}}{\bar \gamma_0 \xi}, \\
&& \cos \phi =  \frac{\left [2 \Delta_0 - D-\xi (4 + 3({\cal
B}_+^2 +{\cal B}_-^2) )\right ]\sqrt{\bar \gamma_+\bar
\gamma_-}}{(\bar \gamma_+ + \bar \gamma_-) \xi}, \nonumber \\  \\
\label{deltp} && \Delta_+ = (2\Delta_0 - D) \frac{\bar
\gamma_+}{\bar \gamma_-+\bar \gamma_+} + \xi (2+{\cal B}_+^2
+{\cal B}_-^2) \times \\ \nonumber && \frac{\bar \gamma_- - \bar \gamma_+}{\bar
\gamma_-+\bar \gamma_+}+ \xi(\bar \gamma_-{\cal B}_-^2 -\bar
\gamma_+{\cal B}_+^2), \\ \label{deltm} && \Delta_- = (2\Delta_0 -
D) \frac{\bar \gamma_-}{\bar \gamma_-+\bar \gamma_+} - \xi
(2+{\cal B}_+^2 +{\cal B}_-^2) \times \\ \nonumber && \frac{\bar \gamma_- - \bar
\gamma_+}{\bar \gamma_-+\bar \gamma_+}- \xi(\bar \gamma_-{\cal
B}_-^2 -\bar \gamma_+{\cal B}_+^2),
\\ && \frac{{\cal B}_+}{{\cal B}_-} = \sqrt{\frac
{\bar \gamma_-}{\bar \gamma_+}} \label{bratio}
\end{eqnarray}
Now, from Eq.~(\ref{f1}) and Eq.~(\ref{f2}) we find equations for $\psi$ and ${\cal B}_-$:
\begin{eqnarray} \label{bamp}
&& \left [ 1+2 \frac{\bar \gamma_-}{\bar \gamma_0} {\cal B}_-^2
\right ]^2 + \\ \nonumber &&
 \biggl [  \Delta_0 - \xi - 2\xi{\cal B}_-^2    \left (
\frac{(\bar \gamma_--\bar \gamma_+)^2}{\bar \gamma_+(\bar
\gamma_++\bar \gamma_-)} + \right.\\ \nonumber && \left. \frac{\bar \gamma_-(2 \Delta_0 -
D)}{\xi (\bar \gamma_++\bar \gamma_-)} -3
\frac{\bar \gamma_-}{\bar \gamma_+} {\cal B}_-^2 \right )\biggr ]^2 = \frac{f^2}{\xi}, \\
&& \cos \psi = \frac{\sqrt \xi}{f} \left [ 1+2 \frac{\bar
\gamma_-}{\bar \gamma_0} {\cal B}_-^2 \right ], \\ \nonumber && \sin \psi =
\frac{\sqrt \xi}{f} \biggl [  \Delta_0 - \xi - 2\xi{\cal B}_-^2
\left ( \frac{(\bar \gamma_--\bar \gamma_+)^2}{\bar \gamma_+(\bar
\gamma_++\bar \gamma_-)} + \right.\\  && \left. \frac{\bar \gamma_-(2 \Delta_0 -
D)}{\xi (\bar \gamma_++\bar \gamma_-)} -3 \frac{\bar
\gamma_-}{\bar \gamma_+} {\cal B}_-^2 \right )\biggr ].
\end{eqnarray}

\subsection{Oscillation frequency}

The complex amplitude of an RF signal generated by the hyper-parametric oscillator output on a fast photodiode is proportional to $A^* B_++A B_-^*$, or
\begin{equation} \label{erf}
E_{RF} \sim |A|^2e^{i \phi_-/2}({\cal B}_+ e^{-i \phi/2}+{\cal B}_- e^{i \phi/2}),
\end{equation}
where $\phi_-=\phi_+-\phi_-$. Equation (\ref{erf}) shows that the phase of the RF signal does not depend on the phase of the pump if ${\cal B}_+={\cal B}_-$. In accordance with Eq.~(\ref{bratio}) this is not true if $\bar \gamma_- \ne \bar \gamma_+$. It means that the phase noise of the pump light will influence the RF phase noise if the modes of the resonator are not identical.

The frequency difference between the pump light and oscillation harmonics can be found from Eqs.~(\ref{deltp}) and (\ref{deltm}) in a straightforward way
\begin{eqnarray} \label{dif}
&& \frac{\tilde \omega_+-\omega }{\bar \gamma_0}=\frac{\omega-\tilde \omega_- }{\bar \gamma_0}= \\ \nonumber
&& \left [ -\frac{D}{2}+\Delta_0-\xi (2+{\cal B}_+^2+{\cal B}_-^2) \right ]\frac{\bar \gamma_--\bar \gamma_+}{\bar \gamma_++\bar \gamma_-}+ \frac{\omega_+-\omega_-}{2\bar \gamma_0}.
\end{eqnarray}
These frequency differences determine the RF frequency  $\omega_{RF}$ produced by the optical signal.  It is possible to rewrite Eq.~(\ref{dif}) in conventional terms as
\begin{eqnarray} \label{freqa}
\omega_{RF}=\omega_{FSR}(T)+ \frac{\Delta Q}{2Q}
\left \{ \frac{\omega_+(T)+\omega_-(T)-2\omega}{2}-   \right. \\ \nonumber \left. g [2|  A   |^2+|  B_+   |^2+|  B_-   |^2] \right \},
\end{eqnarray}
where $\omega_{FSR}(T)$ is the free spectral range of the resonator, $\Delta Q = Q_+-Q_-$ is the difference of loaded quality factors of the two sideband modes, $Q= (Q_{+}+Q_{-})/2$ is the averaged Q-factor of the modes. Equation (\ref{freqa}) means that the frequency of the RF signal generated at the output of optical hyper-parametric oscillator depends both on the detuning of the cw optical pump from the corresponding resonator mode and the power fluctuations of the pump light. Naturally, the noise of these parameters will be transferred to the frequency noise of the RF signal. The transfer coefficient is equal to  $\Delta Q/(2Q)$ and is dependent on the size of the resonator.

If we assume that the loss is given by  Rayleigh scattering, it inversely increases with wavelength as $\lambda^{-4}$ in the vicinity of the carrier frequency of the pump, or
\begin{eqnarray} \label{gamma}
\gamma = \gamma_0 \frac{\omega^4}{\omega_0^4}.
\end{eqnarray}
In accordance with Eq.~(\ref{gamma}) $\Delta Q/(2 Q) \approx -3 \omega_{FSR}/\omega_0$.

The coupling loss also has a dispersion. For instance, in the case of an evanescently coupled WGM resonator, if there is no  gap between the coupling prism and the resonator, the frequency behavior of the resonator loading can be approximated as
\begin{equation} \label{gammac}
\gamma_c= \gamma_{c0}\frac{\omega_0^{3/2}}{\omega^{3/2}}.
\end{equation}
In accordance with Eq.~(\ref{gammac}) we get $\Delta Q_c/(2 Q_c) \approx 5 \omega_{FSR}/(2 \omega_0)$. If a nonzero spatial gap between the coupler and the resonator surface is maintained, the frequency dependence becomes exponential and the coupling efficiency can change by several orders of magnitude if the frequency changes by an octave.

The ratio $\omega_0/\omega_{FSR}$ is equal to the mode order $l$ and is a large number. For instance, $l\simeq 2\times 10^4$ for a resonator with $\omega_{FSR}=2 \pi \times 10^{10}$~rad$/$s pumped at $\lambda_0=1,550$~nm. Therefore, the detuning and laser power noise is suppressed by $4\times 10^8$ times at small frequencies. This value can be reduced if an additional process introducing asymmetry takes place, for example interaction among the modes of the resonator of frequency dependent attenuation resulting from impurities or contaminants (e.g. water).

\section{Numerical simulation of noise transfer from optical pump to frequency comb repetition rate}

The analytical theory of hyper-parametric oscillation presented in the previous section suggests that there is a channel for transfer of  noise from the optical pump to the phase noise of the RF signal generated at the output of the oscillator. This prediction is not accurate enough if we consider generation of the Kerr frequency comb instead of the hyper-parametric oscillation. The frequency comb has many interacting harmonics involved, which can result in modification of the noise transfer efficiency.

To study this particular case we perform a numerical simulation in which  a set of coupled ordinary differential equations is solved to simulate Kerr frequency comb formation \cite{chembo10pra,matsko12pra}. We write the set as
\begin{equation} \label{set}
\dot{a}_j=-(\gamma_j+i\omega_j)a_j+ \frac{i}{\hbar} [\hat V_c,\hat a_j]+F_{j0}(t) e^{-i\omega t}
\delta_{j0,j},
\end{equation}
where $V_c= -\hbar g (e^\dag)^2 e^2/2$ describes nonlinear interaction among modes of the resonator (c.f. Eq.~\ref{v}), $e=
\sum _{j=1}^{j=N} a_j$ is the operator describing amplitude of light confined in the resonator; $a_j$ is an operator standing for complex amplitude of the field in the mode (annihilation operator), $N$ is the total number of modes taken into consideration (in our simulation $N$ does not exceed $141$), $\omega_j$ are  frequencies of the resonator modes, $\delta_{j0,j}$ is the Kronecker's delta. The interaction constant, as defined by Eq.~(\ref{g}), $g=\hbar \omega_{j0}^2 c n_2/({\cal V}n_0^2)$, is assumed to be the same for all the modes; here $\omega_{j0}$ is the frequency of the pumped mode.

Parameters $\omega_j$ carry information about all orders of dispersion of the resonator.
The second order group velocity dispersion $\beta_2$, for example, is related to parameter
$D=(2 \omega_{j0}-\omega_{j0+1}-\omega_{j0-1})/\gamma_{j0} = \beta_2 c\omega_{FSR}^2/(n_0\gamma_{j0})$, introduced above for description of the hyper-parametric oscillator (Eq.~(\ref{dd})), and $2\omega_{FSR} \simeq \omega_{j0+1}-\omega_{j0-1}$. In this paper we do not assume that only second order GVD is present in the resonator, and take the third order dispersion into account to model the resonator spectrum as
\begin{eqnarray}
\omega_j=\omega_{j0}+\omega_{FSR}(j-j_0)-  \frac{D}{2} \gamma_{j0} (j-j_0)^2 + \frac{1}{6}\gamma_{j0} D_3 (j-j_0)^3.
\end{eqnarray}

Similarly to the case of three modes, the half width at  half maximum (HWHM) for the optical modes, $\gamma_j=\gamma_{c\,j}+\gamma_{a\, j}$, consists of two terms, $\gamma_{c\,j}$ and $\gamma_{a\,j}$, standing for coupling and intrinsic loss of the modes. In this paper we consider the case of unequal loss in different modes.

The external optical pumping is given by $F_{j0}(t)= (2 \gamma_{c\,j0} P(t)/(\hbar
\omega_{j0}))^{1/2} \exp [i\phi_{j0}(t)]$, where $P_{j0}(t)$ is the value of the power of the pump light and
$\phi_{j0}(t)$ is time dependent phase of the pump light.

To find the transfer function characterizing the impact of the optical noise on the repetition rate of the
Kerr frequency comb we introduce slow modulation of the pump light and find the relative power of modulation
of the RF signal generated by the frequency comb. Namely, we present the
normalized external force in the form
\begin{eqnarray} \label{modulation}
f_{j0}(t)= F  \left
[1+(i\kappa_\phi+\kappa_a )\cos(2 \pi f_0 t) \right ], \\
F=\sqrt{\frac{2  P_{j0}}{\hbar \omega_{j0} \gamma_{j0}} \frac{g}{\gamma_{j0}} },
\end{eqnarray}
where $\kappa_\phi$ and $\kappa_a$ are coefficients of phase and amplitude modulation, $f_0$ is the
modulation frequency. We assume that the modulation coefficients are much less than unity since we are interested in the study of transfer of a relatively small optical noise to the RF signal.

\subsection{Examples of frequency combs investigated}

We find the region of solutions describing fundamentally mode locked frequency combs by solving Eqs.~(\ref{set}) numerically. Those combs correspond to generation of a single optical pulse within the resonator.  This pulse generates trains of freely propagating optical pulses by interacting with the input/output couplers during each round trip. The loss associated with the train generation is compensated by interaction of the intracavity pulse with the continuous wave background powered by the external pump.

The pulse duration, $\tau$, and corresponding spectral width of the frequency comb, $\Delta f_{comb}$, can be approximated analytically
\begin{eqnarray} \label{tau}
\tau \approx  \sqrt{\frac{-2 \beta_2 \lambda_0 A_{eff}}{\pi^2 {\cal F} P_{in} n_2}}, \\
\Delta f_{comb} \simeq (\pi^2 \tau)^{-1}, \label{df}
\end{eqnarray}
where $\lambda_0$ is the wavelength of the pump light \cite{matsko13oe,coen13ol1,herr14np}. As a rule, the smaller  the GVD ($\beta_2$), the broader is the comb for the same  set of parameters. We selected two cases of comparably large and comparably small GVD to study optical-to-RF noise transfer for the signal generated by the light produced by the hyper-parametric oscillation, and mode locked Kerr frequency comb.

\begin{figure}[ht]
  \centering
  \includegraphics[width=6.0cm]{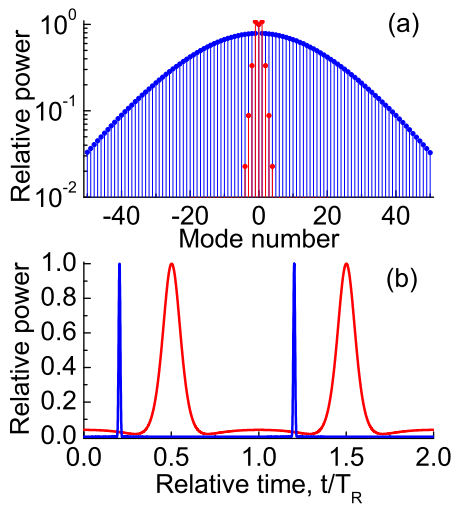}
\caption{ \small Spectra of two mode locked optical frequency combs, panel (a), used in the simulation and
corresponding optical pulses, panel (b), generated in the resonator. The narrow frequency comb is generated in a
resonator with GVD $D=-2$ by cw pump $F=2.5$ and pump carrier frequency detuned from the resonance by $\Delta_0=4.1$. Only the first two sidebands are pronounced in the frequency comb, so it can be considered as a well developed hyper-parametric oscillation. A wide frequency comb is generated for $D=-0.05$ by pump $F=7.1$ and pump frequency detuned from the resonance by $\Delta_0=44$. This is a well developed Kerr frequency comb. } \label{fig2}
\end{figure}

The envelopes of the optical frequency combs we used in the simulation are shown in Fig.~(\ref{fig2}a). While the Kerr frequency comb corresponds to generation of short optical pulses, the hyper-parametric oscillator naturally forms pulses that rather resemble a beat note of several mutually coherent harmonics (Fig.~\ref{fig2}b). Generation of the combs in both cases have hard excitation regime \cite{matsko12pra}. To simulate them numerically we select an arbitrary nonzero initial conditions for the field localized in the resonator modes. Practically the combs can be generated by proper nonadiabatic changing of the pump frequency and$/$or power \cite{matsko13ol}.

\subsection{Simulation technique}

Once the frequency and the power are selected to observe the frequency comb (Fig.~\ref{fig2}), we introduce modulation on the cw pump per Eq.~(\ref{modulation}). The pulse trains are generated at both add (input) and drop (output) ports of the nonlinear resonator (Fig.~\ref{fig1}). The modulation coefficient is selected to be much less than  unity to ensure that the generated frequency comb stays in the same regime with and without the modulation.

We select the pulse train exiting the drop port and define the phase of the RF signal generated by the comb as
\begin{equation} \label{psi}
\Psi(t)= {\rm Arg} \left [ e^{ i (\omega_{FSR}+ 2\pi \delta)\ t} \sum \limits_j {\hat a}_j {\hat a}_{j-1}^\dag \right ],
\end{equation}
where $\delta$ corresponds to the deviation of the comb repetition rate from the FSR, occurring due to the presence of the wavelength dependent Q-factor, as well as high order dispersion. Similarly, we define the power of the frequency comb as
\begin{equation} \label{pcomb}
P_{comb}(t)=  \sum \limits_j {\hat a}_j^\dag {\hat a}_{j}.
\end{equation}
The power of the frequency comb is proportional to the amplitude of the base-band RF signal generated by the comb on a fast photodiode.

\begin{figure}[ht]
  \centering
  \includegraphics[width=6.0cm]{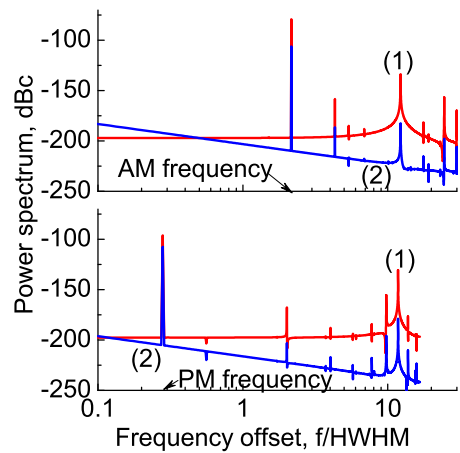}
\caption{ \small  Examples of simulated power spectra $S(f)$ of the RF signal. Curve (1) stands for a narrow frequency comb,
curve (2) stands for the broad frequency comb (see Fig~\ref{fig2}a).  Top panel stands for the amplitude modulation, bottom one -- for phase modulation of the pump. The modulation frequency is shown by the arrows. The resonator has $D=-0.05$ and $D_3=-10^{-4}$ to achieve the noise transfer. The AM and PM modulation coefficients ($\kappa_a$ and $\kappa_\phi$) are selected to be the same, $10^{-3}$. }
\label{fig3}
\end{figure}

The simulation is performed over hundreds of ring-down times of the resonator, to ensure that the system reaches its steady state.  The phase and power of the comb, $\Psi(t)$ and $P_{comb}(t)$, are recorded. Then we use a fast Fourier transform (FFT) to evaluate the power spectrum of these parameters (Fig.~\ref{fig3}) and find the maximum of the harmonic corresponding to the modulation frequency. The transfer function we are looking for is represented by the frequency dependence of the modulation harmonic. We define the transfer functions as
\begin{eqnarray} \label{tr1}
H_{PM}(f)= \frac{4}{\kappa_{\phi}^2} S_{PM\ max}(f), \\
H_{AM}(f)= \frac{4}{P_{comb} \kappa_{a}^2} S_{AM\ max}(f). \label{tr2}
\end{eqnarray}
where $S_{PM\ max}(f)$ and $S_{AM\ max}(f)$ are the maxima of the spectral power densities of the RF phase, Eq.~(\ref{psi}), and comb power, Eq.~(\ref{pcomb}), signals, respectively.

We found that the transfer functions is zero (their values are smaller than the digital accuracy of the simulation) when an ideal resonator with identical modes and only the second order GVD is considered. However, when higher order dispersive terms and$/$or frequency dependence of the resonator Q-factor are introduced, the transfer functions start to deviate from zero. This means that, in accordance with the simplified analytical treatment presented above, the modulation of the pump light can be transferred to the frequency comb parameters. Since small additive noise can be considered as modulation of a regular signal with a small noise signal, the transfer functions can be used to evaluate the noise transfer from the cw pump light to the frequency comb parameters.

\subsection{Conversion of optical amplitude noise to frequency comb noise due to frequency dependent Q-factor}

Quality factors of modes of any realistic resonator are not identical. In WGM resonators, for instance, there are fundamental reasons that enforce the inequality of Q-factors of any two adjacent optical modes. This inequality results in fundamental mechanism for noise transfer from the pump light to the generated pulse train and to the corresponding optical frequency comb.

\begin{figure}[ht]
  \centering
  \includegraphics[width=6.0cm]{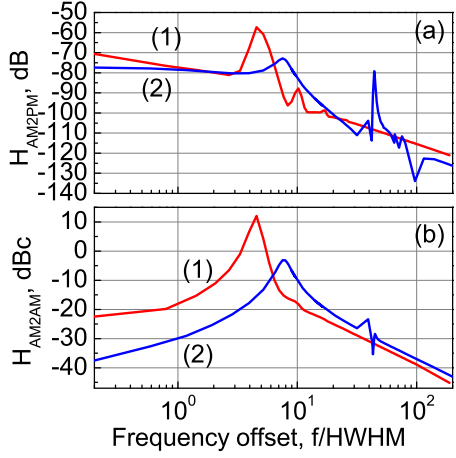}
\caption{ \small Transfer functions characterizing the impact of amplitude modulation (amplitude noise) of the pump light on the frequency comb, simulated for the case of frequency dependent loss of the resonator modes: $\gamma_{j}/\gamma_{j0}=1 + 10^{-4}(j - j_0)$. The rest of the parameters are the same as in Figs.~\ref{fig2} and \ref{fig3}. Panel (a) describes AM to PM conversion in  narrow and broad frequency combs, while panel (b) stands for AM to AM conversion. Red lines, (1), stand for hyper-parametric (narrow) oscillation comb, blue lines, (2), stand for the (broad) Kerr frequency comb. } \label{fig4}
\end{figure}

We introduce the frequency dependence of the quality factors and evaluate the transfer functions using Eqs.~(\ref{tr1}) and (\ref{tr2}). The result of the simulation is shown in Fig.~(\ref{fig4}). The amplitude modulation impacts both the phase and the amplitude of generated frequency combs. There are two well pronounced resonances in the transfer functions. The resonances occur at frequency $f_1 \approx 9 HWHM$ and $f_2 \simeq \Delta_0$. The low frequency, $f_1$, resonance corresponds to the relaxation oscillation. It depends on the Q-factor of the resonator and less on the other parameters of the system. The high frequency, $f_2$ resonance corresponds to the detuning of the pump frequency from the frequency of the cold resonator mode. This resonance has an analogy with Feshbach resonances in atomic systems \cite{matsko14arch}.

The numerical simulation shows that the  hyper-parametric oscillator has higher efficiency for optical-to-RF modulation transfer. It also predicts that the amplitude modulation of the pump light can be transferred with amplification to the power of the frequency comb if the modulation frequency corresponds with the relaxation oscillation resonance. It means that the amplitude noise of the pump laser can be efficiently imprinted on the amplitude noise of the RF signal generated by the optical frequency comb on a fast photodiode.

\subsection{Conversion of optical phase noise to frequency comb noise due to frequency dependent Q-factor}

\begin{figure}[ht]
  \centering
  \includegraphics[width=6.0cm]{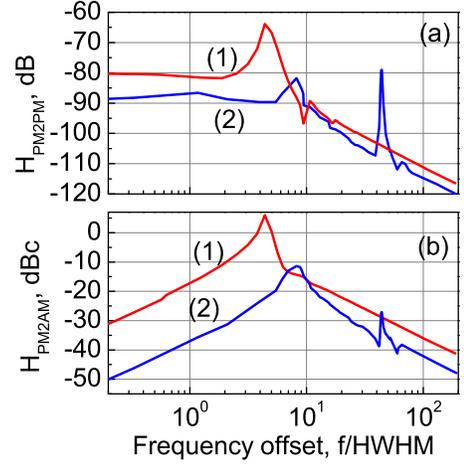}
\caption{ \small Transfer functions characterizing the impact of phase modulation (phase noise) of the pump light on the frequency comb simulated for the case of frequency dependent loss of the resonator modes. The parameters of the system are selected in the same way as in Figs.~\ref{fig2}--\ref{fig4}. Panels (a) and (b) correspond to PM to PM conversion and  PM to AM conversion, respectively. Red lines, (1), stand for the hyper-parametric oscillation comb, blue lines, (2), stand for the broad Kerr frequency comb.} \label{fig5}
\end{figure}

We repeated the simulation for phase modulated pump light using identical parameters of the system. The results of the simulation are shown in Fig.~(\ref{fig5}). They are similar to those obtained for the amplitude modulated pump. Phase modulation (phase noise) of the pump light impacts both amplitude and phase of the generated frequency combs (power and timing of the corresponding pulse train).

\subsection{Conversion of optical noise to frequency comb noise due to the presence of third order dispersion}

We found that frequency dependent Q-factor of the spectrum of a nonlinear resonator initiates transfer of modulation (noise) of the pump light to the parameters of generated optical frequency comb. There is another parameter of the spectrum that results in the noise transfer. This parameter is related to high-order dispersion of the resonator modes. The high order dispersion is always present in the resonator due to dispersion of the resonator host material, which has nonlinear dispersion due to limited transparency bandwidth. The wavelength-dependent quality factor of the resonator spectrum also has to have certain correspondence with the dispersion, in accordance with the Kramers-Kronig relationship.

\begin{figure}[ht]
  \centering
  \includegraphics[width=6.0cm]{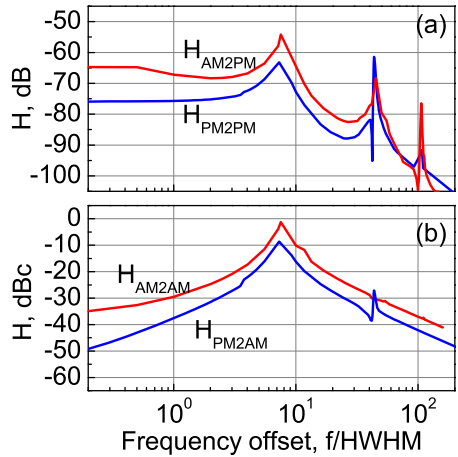}
\caption{ \small  Transfer functions describing impact of the pump modulation on the phase (panel (a)) and amplitude (panel (b)) of the Kerr frequency comb for $D = -0.05$ and $D_3 = -0.0001$. The other parameters of the system are selected in the same way as in Figs.~\ref{fig2} and \ref{fig3}.
} \label{fig6}
\end{figure}

For the sake of clarity we neglect the wavelength dependence of the absorption and introduce a small third order dispersion (TOD) to the resonator spectrum. The numerical simulations performed with this kind of resonator clearly show that the presence of a small TOD results in noise transfer similar to what we observed with frequency dependent Q-factor. We present results of the simulation for the (broad) Kerr frequency comb only (Fig.~\ref{fig6}), since the results for hyper-parametric oscillation have similar behavior, as shown for the frequency dependent Q-factor.

\subsection{Impact of limited mode number on simulation result}

\begin{figure}[ht]
  \centering
  \includegraphics[width=7.0cm]{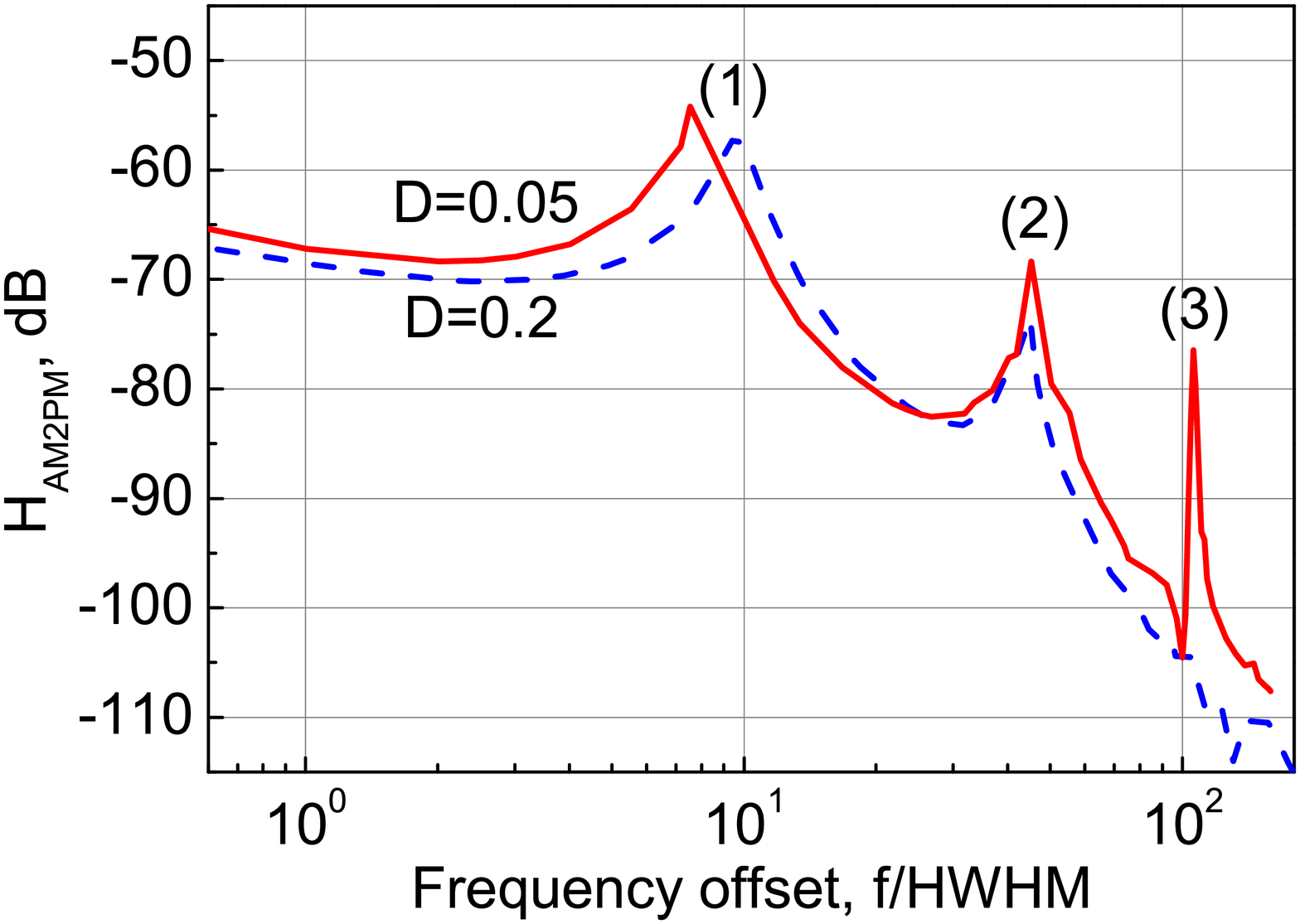}
\caption{ \small The result of numerical simulation showing that the resonance (3) depends on GVD of the resonator spectrum. The dashed blue line corresponds to $D=-0.2$, while sthe olid red line stands for  $D=-0.05$. The other conditions are the same for both curves. Resonance (1) corresponds to relaxation oscillation of the system and resonance (2) is the Feshbach resonance. } \label{fig7}
\end{figure}

Similarly to the case of wavelength dependent Q-factor we observed two resonances at the transfer functions shown in Fig.~(\ref{fig6}). One of the resonances corresponds to the relaxation oscillation and the other -- to a Feshbach resonance. We also observed a third resonance in the $H_{AM2PM}$ curve (Eq.~\ref{fig6}). To understand the nature of the resonance we repeated the simulation with different second-order GVD parameters. We found that the resonance shifts beyond the dynamic range of the simulation accuracy (practically disappears) for larger GVD values (when we change $D=-0.05$ to $D=-0.2$); the relaxation oscillation resonance shifts to smaller frequencies, and the Feshbach resonance changes its contrast, not the frequency, by a small amount (Fig.~\ref{fig7}). The third resonance also disappears if we change the number of modes in the simulation. These observations show that the resonance is a consequence of the limited mode number we have considered. Physically, the resonance can be interpreted as a feature occurring due to the presence of an additional filter function in the system, because the limited number of modes under consideration is equivalent to the presence of a square band-pass filter inserted in the nonlinear ring resonator.

\section{Discussion}

Our simulations show that while transfer of the optical pump amplitude noise to the noise of the fundamentally mode locked frequency comb is present in realistic nonlinear resonators, it is rather small. For instance, if a pump laser has RIN of $-140$~dB$/$Hz at 10~kHz and the value of the AM-to-PM transfer function is -75~dB, the noise transfer can be safely neglected, since the best RF phase noise realized with a Kerr frequency comb oscillator approaches $-130$~dBc$/$Hz at 10~kHz \cite{liang15tbp}; that is much larger than -225~dBc/Hz value resulting from the noise transfer. The same conclusion applies to higher offset frequencies.

The impact of the laser phase noise, however, can be more significant. For instance, if the phase noise of the pump laser is $-50$~dBc$/$Hz at 10~kHz, which is a realistic value for a stabilized diode laser, and the PM-to-PM transfer function value is -80~dB at 10~kHz, the associated phase noise of the RF oscillator is limited by -130~dBc/Hz at 10~kHz. This number is already significant for high-quality RF photonic oscillators.

The situation can become even worse if modes of the resonator interact. It was shown that mode interaction changes the frequency spacing (GVD) of the resonator modes rather significantly \cite{savchenkov12oe,herr13arch,liu14arch,xue14arch,ramelow14ol}. The change can be so large that fundamental hyper-parametric oscillations with soft excitation take place in a resonator with small normal GVD. The mode interaction also destroys the symmetry of the mode Q-factors, leading to several percent differences among Q-factors of adjacent modes. It may result in five to six orders of magnitude increase in the noise transfer efficiency for both amplitude and phase noise.

The highest effect of the noise transfer is expected in the vicinity of the relaxation oscillation peak (the phase noise at smaller frequency is usually determined by thermal fluctuations of the resonator). If the phase noise arising from the detuning frequency noise of the laser and the resonator mode is -50~dBc/Hz at 10~kHz and the PM-to-PM transfer function flat part reaches -40~dB, the resultant RF phase noise will be at the level of -90~dBc/Hz, which is a large number for, e.g., an RF Kerr comb oscillator.

Another possibility for increased efficiency of the noise transfer occurs in multi-soliton combs, which have certain analogy with harmonic mode locking. Those combs correspond to multiple optical pulses confined in the nonlinear resonator. In such a regime the power of the RF frequency signal generated by the comb on a fast photodiode is higher than in the case of the fundamentally mode locked comb. On the other hand, it was found that the sensitivity of the multi soliton combs to the modulation of the pump light is several orders of magnitude stronger than the sensitivity of the fundamentally mode locked comb discussed in this paper \cite{matsko14arch}.

Therefore, to achieve highly spectrally pure RF signals generated in a photonic oscillator based on a Kerr frequency comb, one needs to operate the device in the fundamentally mode locked regime and also use a pump laser with as low amplitude and phase noise as possible.

\section{Conclusion}

We have analyzed the impact of the noise of continuous wave pump light on the noise of the RF signal generated by a Kerr frequency comb based photonic oscillator. Using both analytical and numerical simulations we have shown that the optical noise can be transferred to the comb noise in the case where the resonator spectrum has a certain asymmetry, such as nonzero third- and higher order dispersion and frequency-dependent quality factor. The noise transfer can already be significant at the current level of the technology  and future experimental research is needed to further explore the impact of the optical pump noise in Kerr frequency comb oscillators.

\section*{Acknowledgment}
The authors acknowledge support from Defense Sciences Office of Defense Advanced Research Projects Agency under contract No. W911QX-12-C-0067 as well as support from Air Force Office of Scientific Research under contract No. FA9550-12-C-0068.


\end{document}